\documentclass[pteplogo]{ptephy_v2}
\usepackage[normalem]{ulem}

\begin{document}


\title{Removal of Radioactive Noble Gas Radon from Air by Ag-Zeolite}

\author[1]{H.~Ogawa}
\affil[1]{CST Nihon University, Surugadai, Kanda, Chiyoda-ku, Tokyo, 180-0011, Japan. \email{ogawa.hiroshi@nihon-u.ac.jp}}
\author[2,3]{Y.~Takeuchi} 
\affil[2]{1Department of Physics, Graduate School of Science, Kobe University, Kobe, Hyogo
657-8501, Japan}
\affil[3]{Kavli Institute for the Physics and Mathematics of the Universe (WPI), The University of
Tokyo Institutes for Advanced Study, University of Tokyo, Kashiwa, Chiba 277-8583, Japan.}
\author[3,4]{H.~Sekiya}
\affil[4]{Kamioka Observatory, Institute for Cosmic Ray Research, University of Tokyo, Kamioka, Gifu 506-1205, Japan}
\author[5]{K. Iyoki}
\affil[5]{Department of Environment Systems, Graduate School of Frontier Sciences, The University of Tokyo, Kashiwanoha, Kashiwa-shi, Chiba 277-8563, Japan}
\author[6]{M. Matsukura}
\affil[6]{Institute of Engineering Innovation, The University of Tokyo, Yayoi, Bunkyo-ku, Tokyo, 113-8656, Japan}
\author[6,7]{T. Wakihara}
\affil[7]{Department of Chemical System Engineering, The University of Tokyo, Hongo, Bunkyo-ku, Tokyo, 113-8656, Japan}
\author[8]{Y. Nakano}
\affil[8]{Faculty of Science, University of Toyama, Gofuku, Toyama-shi, Toyama 930-8555, Japan}
\author[9]{S. Hirano}
\affil[9]{Tosoh Corporation, Tokyo Midtown Yaesu, Yaesu Central Tower, Yaesu, Chuo-ku, Tokyo 104-8467 Japan}
\author[10]{A. Taniguchi}
\affil[10]{Sinanen Zeomic Co., Ltd, Nakagawa-honmachi, Minato-ku, Nagoya City, Aichi Prefecture, Japan}

\newcommand{\revi}[2]{\textcolor{red}{\sout{#1}}\textcolor{blue}{#2}}
\newcommand{\add}[1]{\textcolor{blue}{#1}}

\begin{abstract}
This paper investigates the removal of radon from purified and ambient airs by Ag-zeolite. Ag-zeolite is known to have very high performance for airborne radon removal. 
The dependence of zeolite type and silver content on the performance of radon removal was evaluated. The performance of radon removal by single pass and radon emanation were also evaluated. In addition, the adsorption performance due to zeolite solidification and the change in adsorption performance due to moisture in the gas were evaluated in order to investigate properties that should be considered for future practical use.
These properties will be adaptable to the development of air purification systems for the large volume ultra-low radioactivity experiments.

\end{abstract}
\subjectindex{xxxx, xxx} 
\maketitle



\section{Introduction}
\label{sec:intro}
Neutrino experiments and direct dark matter search experiments require an extremely low radioactivity environment at the facility where the experimental detectors are constructed~\cite{Ianni}. In particular, radioactive radon gas in the air must be reduced to the utmost limit because it generates radioactivity from the rocks in the underground facility and the detector itself. Radon is produced from radioactive impurities of the uranium and thorium series in members and bedrock. Radiation from the beta decay of $^{214}$Bi, the daughter nucleus of $^{222}$Rn, puts a limit on the observed energy threshold for solar neutrino observations in water Cherenkov detector, such as Super-Kamiokande~\cite{NIM-SK, SK-solar, Fukuda}. In addition, beta radiation from $^{214}$Pb which is the daughter nucleus of $^{222}$Rn, and nuclear recoil from alpha decay of $^{210}$Po, which attaches to the detector surface, can be background events in low energy solar neutrino and dark matter search experiments.

Airborne radon removal has been done at these extremely low radioactivity experimental facilities. But in recent years, as experimental facilities have become larger, the development of more efficient radon removal equipment has been demanded. For example, the Hyper-Kamiokande water Cherenkov detector under construction in Hida City, Gifu Prefecture, Japan~\cite{HKdesign}, is larger size than the Super-Kamiokande, requiring 3 times the processing capacity of the air purification system toward 1~mBq$/$m$^{3}$ radon contcentration that used to use activated carbon~\cite{SKair, Nakano}.

A promising adsorbent with even better radon adsorption capacity is silver (Ag) ion exchanged zeolite (Ag-zeolite). Ag-zeolite can be prepared by an ion-exchange process that introduces Ag ions into the zeolite pore structure. Gases are adsorbed by chemisorption through the interaction of Ag ions and electrons of gas. Adsorption technology  using Ag-zeolite has been studied extensively in recent years~\cite{Fukui}. 
In addition, Ag-zeolite was found to have excellent radon adsorption performance. Ag-zeolite retains its radon adsorption capacity even at room temperature~\cite{SHeinitz, Oleksandra}. Therefore, it is expected to develop an air purification system that is less expensive, compact, and can be operated at room temperature. The use of Ag-zeolite as a purification device for ultra-low radioactivity experiments is a new idea that has not yet been put to practical use.

In this paper, the sample of Ag-zeolites were fabricated and their adsorption performance on radon in air was investigated. Two types of based zeolites and samples with different exchange ratios of Ag ions were investigated as Ag-zeolites. The emanation of radon, the effect of moisture and the effect of solidification were also investigated. Then, prospects for the development of a new air purification system are discussed.

\section{The synthesis of the Ag-zeolite}
\label{sec:development}
\begin{figure}[htbp]
  \begin{center}
    \includegraphics[keepaspectratio=true,height=50mm]{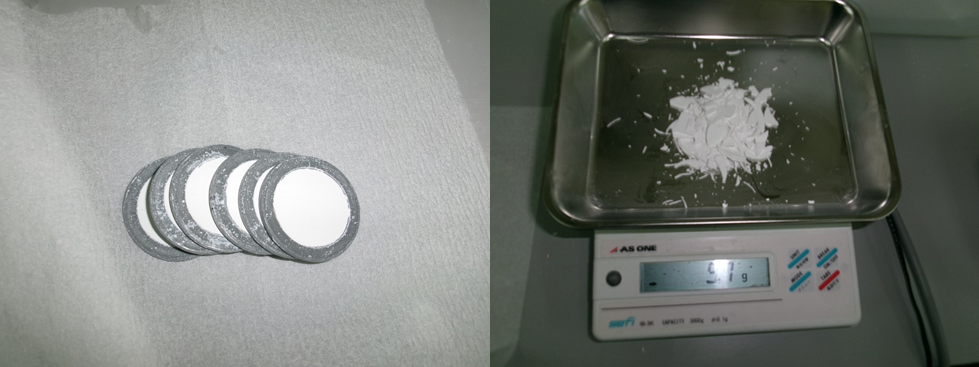}
  \end{center}
  \caption{Photographs of the zeolite sample Ag-MFI. (Left) Solidified sample placed in mold. (Right) A sample being removed from a mold and weighed on a scale.}
  \label{pict:sample}
\end{figure}

Three different Ag-zeolite samples were prepared: two base zeolites, Zeolite Socony Mobil–5~(MFI type zeolite, SiO$_{2}$/Al$_{2}$O$_{3}$$\sim$40~(mol/mol)) and  Ferrierite (FER type zeolite SiO$_{2}$/Al$_{2}$O$_{3}$$\sim$18~(mol/mol)). We selected these zeolites because the aluminum content (Al/Si) in the zeolite that is actually ion-compound with silver is relatively large. These bases were Na form and were ion-exchabged to Ag in AgCl solution. In particular, for FER, we prepared samples with Ag loading amount of 3.0 wt$\%$ (Ag-FER3$\%$) and 7.8 wt$\%$ (Ag-FER8$\%$). The base zeolite was used as is a powder purchased from Tosoh Co., and the Ag ion exchange was performed by Sinanen Zeomic Co., Ltd. Furthermore, these samples (Ag-MFI with 3.3 wt$\%$ for Ag loading amout, Ag-FER3$\%$ and Ag-FER8$\%$) were pressure solidified by Sinanen Zeomic Co., Ltd. Figure~\ref{pict:sample} shows the picture of the Ag-MFI sample. These samples were synthesized in June 2023. The amount of samples synthesized is 20 g each. In addition, a sample of zeolite MFI without ion exchange was also prepared.

\section{Performance of synthesized Ag-zeolite in gases}
\label{sec:perform}
\subsection{The radiactivity measurement system and filter housing}
\begin{figure}[htbp]
  \begin{center}
    \includegraphics[keepaspectratio=true,height=80mm]{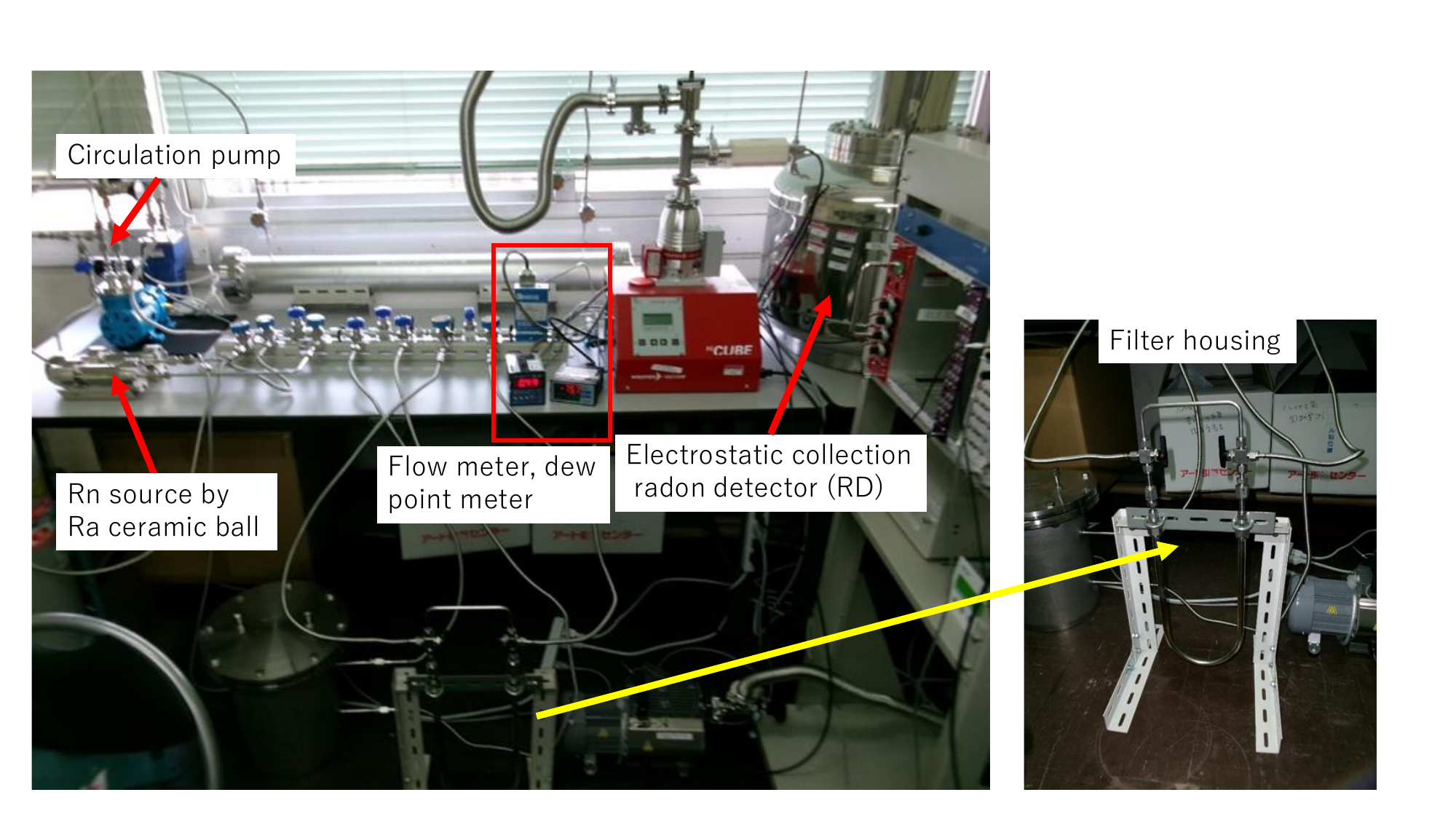}
  \end{center}
  \caption{An overview photograph of the radioactivity measurement system and FH. The line setup is closed system in this photograph.}
  \label{fig:setup}
\end{figure}

For evaluating its intrinsic radioactivity, we used the radioactivity measurement system, that is described in detail in Ref.~\cite{Ogawa1}.  Figure~\ref{fig:setup} shows the overview of the radioactivity measurement system for this study. The 80~L volume radon detector (RD) with an electrostatic collection was used for the estimation of radon concentration in air, which measures the Rn concentration in gas by detecting alpha-ray from $^{214}$Po, the daughter nucleus of $^{222}$Rn~\cite{Hosokawa}. The measurement data is recorded in a multi-channel analyzer with Rasberry Pi. A air circulation pump (Enomoto Micro Pump Mfg. Co., Ltd., MX-808ST-S) was used to circulate the air. Radium ceramic balls are available as a $^{222}$Rn sources, and radon can be introduced into the circulated air in advance for radon removal tests. A flow controller (KOFLOC Co. MODEL 5410, 10$\%$ accuracy at 0.5~L/min), and a dew point meter (TEKHNE Co. TH-100, $\pm$2$^{\circ}$C accuracy) to monitor the moisture in the circulating air are used. 

The prepared zeolite was crushed and placed in a filter housing made of a bent stainless pipe, half-inch in diameter and 25 inches in length as shown in Fig.~\ref{fig:setup} by following procedure. The actual zeolite-filled length is about 12 inches. The filter housing (FH) has a metal nanoparticle filter (NASclean{\textregistered} GF-T001) in both ends. The FH was activated before the test by heating with a ribbon heater at 350$^{\circ}$C for six hours with the evacuation from the both side of the FH. In the Rn removal test of this study, FH is performed in a room temperature environment adjusted to 25$^{\circ}$C. 
Here we introduce two flow conditions for two measurements. One is that purified air was circulated between filter system and RD as a ``closed system'' for evaluating the retainsion time of zeolite. The other is that  the room air was injected from filter system to RD as ``opened system'' for evaluating the radon removal efficiency under the  one-pass situation, which is similar situation with actual air purification system for ambient air. These systems will be explained in each following sections.
  
\subsection{Radon removal measurement by closed system}
\label{subsec:close}
\begin{figure}[htbp]
  \begin{center}
    \includegraphics[keepaspectratio=true,height=70mm]{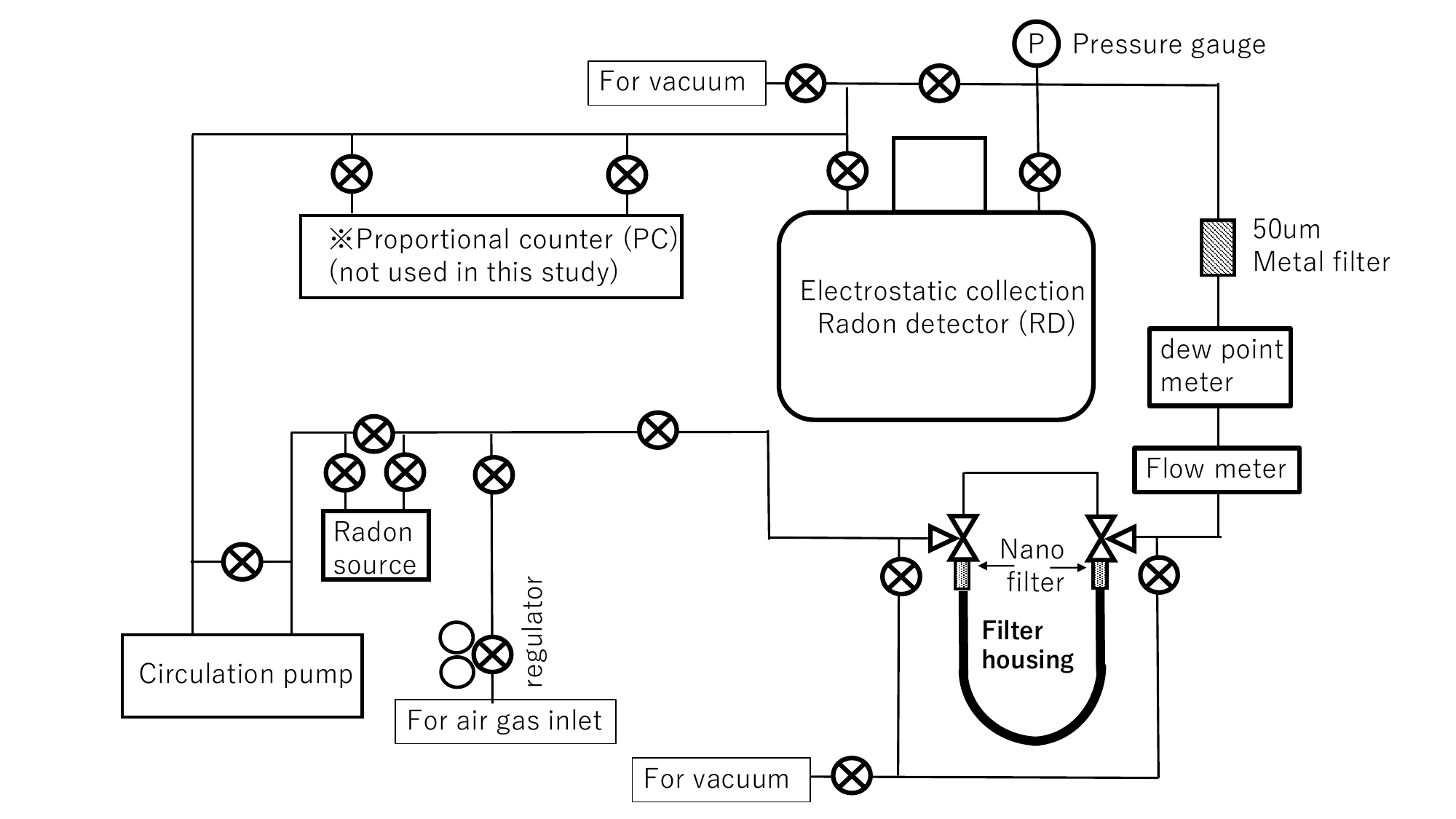}
 \end{center}
  \caption{A schematic representation of the gas line and the radioactivity measurement system for closed system. }
  \label{fig:closed}
\end{figure}

The radon removal measurements were conducted with closed system as follows. Figure~\ref{fig:closed} shows a schematic representation of the gas line and the radioactivity measurement system for closed system. 
Once the radon detector and the entire line were evacuated, commercially-available G1-grade high-purity air (impurity concentration~$<$~0.1~ppm) was filled into the whole system at 1 atm. The circulation of air was then started at 0.5 L/min. Radon was injected from the radon source by passing air gas. After bypassing the radon source, air gas was circulated about 1 day to ensure a stable radon count rate, estimated by counting the decay of $^{214}$Po alpha-ray. Then, the circulating air passes through the filter (filter ON). 

\begin{figure}[htbp]
  \begin{center}
    \includegraphics[keepaspectratio=true,height=60mm]{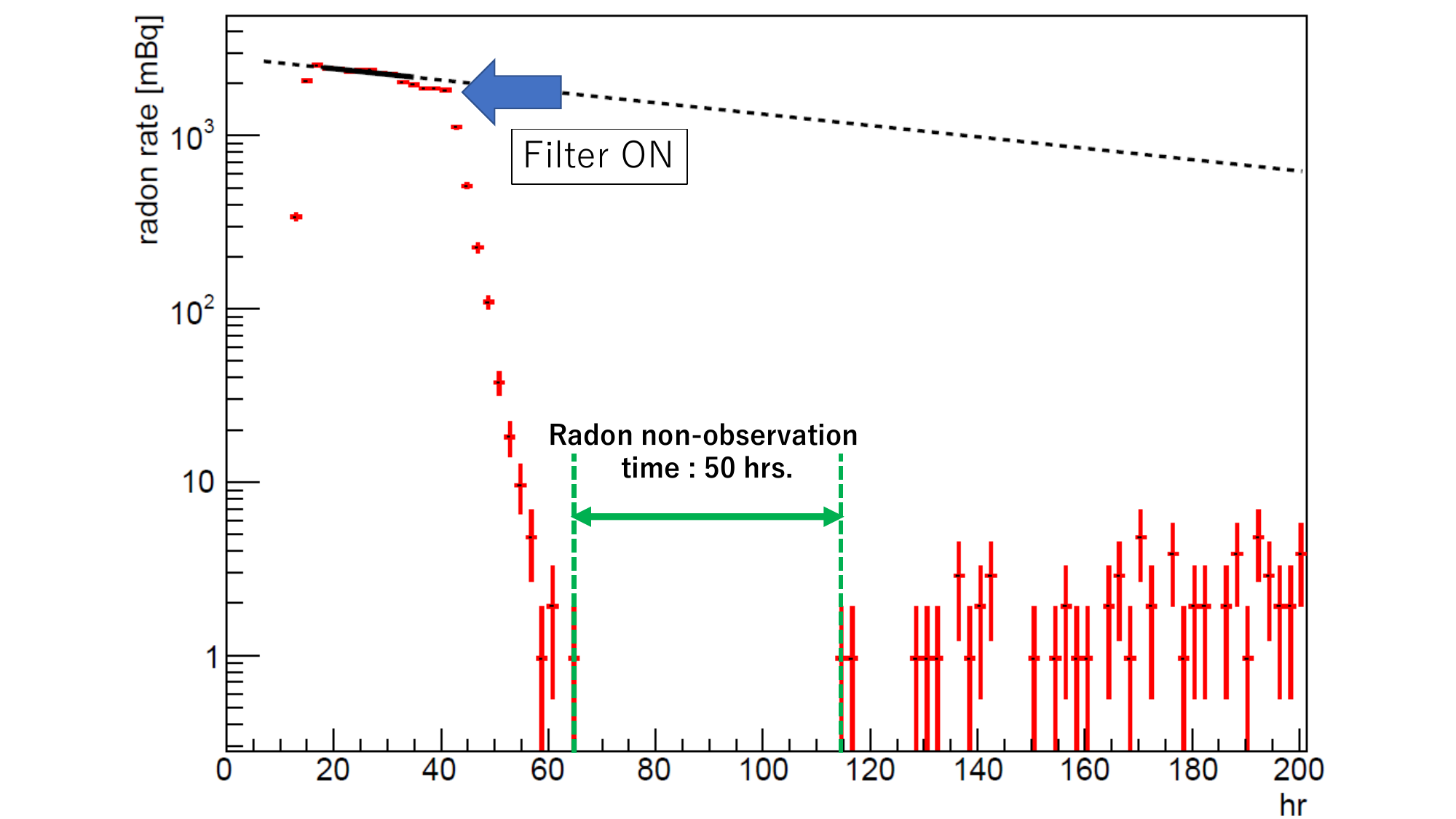}
  \end{center}
  \caption{The time profile of the radon counting rate during the radon removal test with Ag-MFI sample.}
  \label{fig:profile}
\end{figure}

Figure~\ref{fig:profile} shows the time profile of the radon count rate in unit of mBq during the radon removal test with Ag-MFI sample. The radon counting rate was decreased suddenly after filter ON then no radon signal is observed during 50 hours. Then the radon rate is slightly recovered after 50 hours. 

\begin{figure}[htbp]
  \begin{center}
    \includegraphics[keepaspectratio=true,height=80mm]{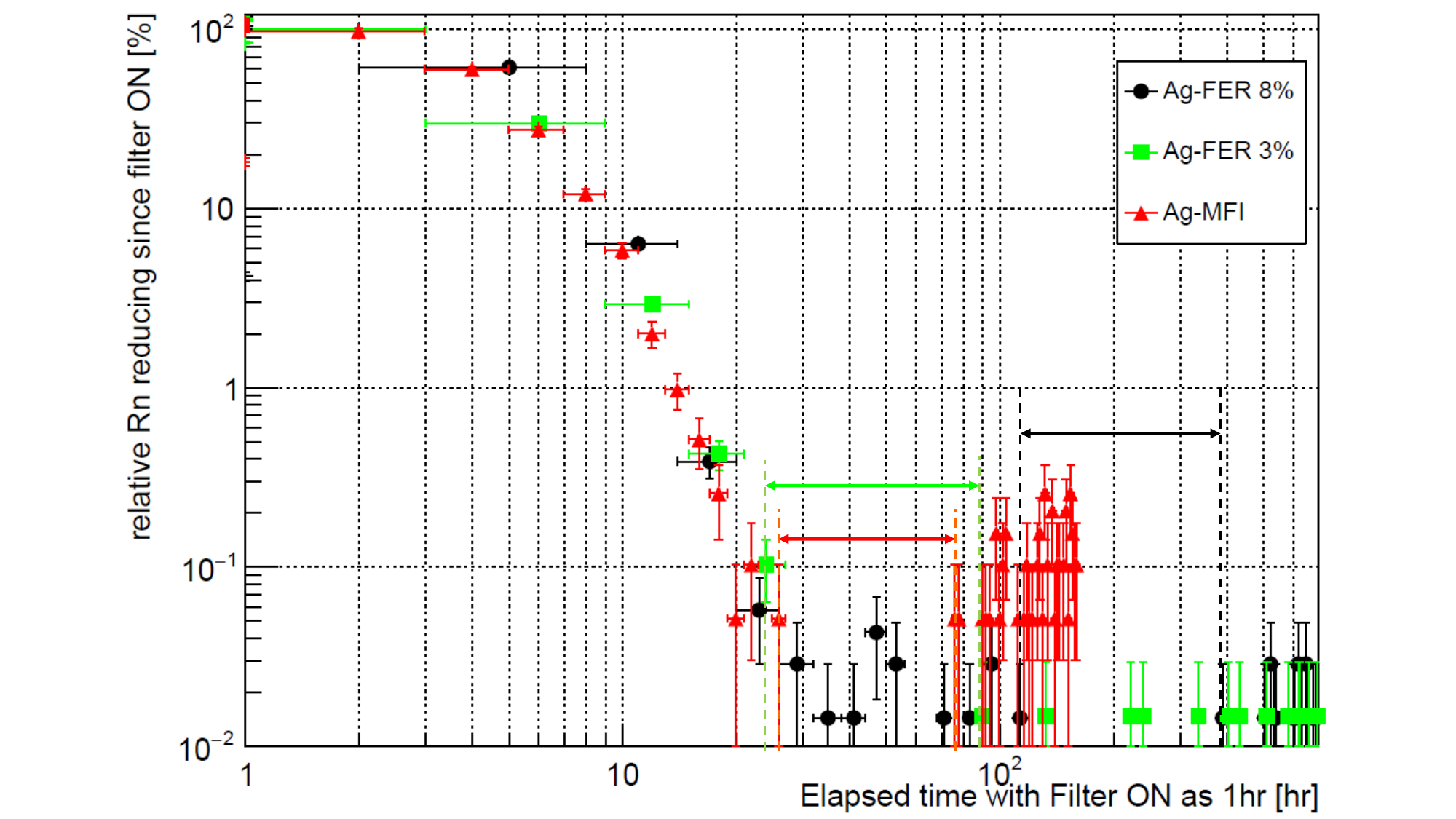}
  \end{center}
  \caption{The time profiles of radon reduction factor in the three radon removal tests. black-circle, green-square, and red-triangle indicate the results for Ag-FER8$\%$, Ag-FER3$\%$ and Ag-MFI, respectively. Each interval and arrow indicates the RnNO time interval in each sample defined in the text.}
  \label{fig:compare}
\end{figure}

We compared the radon removal efficiency of three samples.
Figure~\ref{fig:compare} presents the time profiles of radon reduction factor in the three radon removal tests. The three samples tested in this study showed order 10$^{4}$ radon removal at room temperature. This result demonstrates that zeolite has a high radon removal capacity. Figure~\ref{fig:compare} also shows the radon non-observation time (RnNO time) for each sample.  RnNO time is defined as the time from the disappearance of radon to the appearance of a small amount of radon.
Table~1 shows the measurement conditions for each of the three samples and the respective RnNO times. 
Even though their mass is similar, Ag-FER8$\%$ was found to have the longest RnNO time. This indicates that the adsorption rate of radon depends on the Ag content.


\begin{table}[htbp]
\begin{center}
\caption{The measurement conditions for each of the three samples and the respective RnNO times in the closed system.}
\begin{tabular}{lcccc}
    \hline \hline
    sample&Ag loading amout&sample weight&flow rate&RnNO time\\
   name&[wt$\%$]&[g]&[L/min]&[hours]\\
    \hline
    Ag-MFI&3.3&20.9&0.5&50\\
   Ag-FER3$\%$&3.0&20.2&0.5&66\\
    Ag-FER8$\%$&7.8&20.3&0.5&276\\
    \hline \hline
\end{tabular}
\end{center}
\label{tabsum}
\end{table}
\subsection{Radon removal measurement by open system}
\label{subsec:open}
\begin{figure}[htbp]
  \begin{center}
    \includegraphics[keepaspectratio=true,height=70mm]{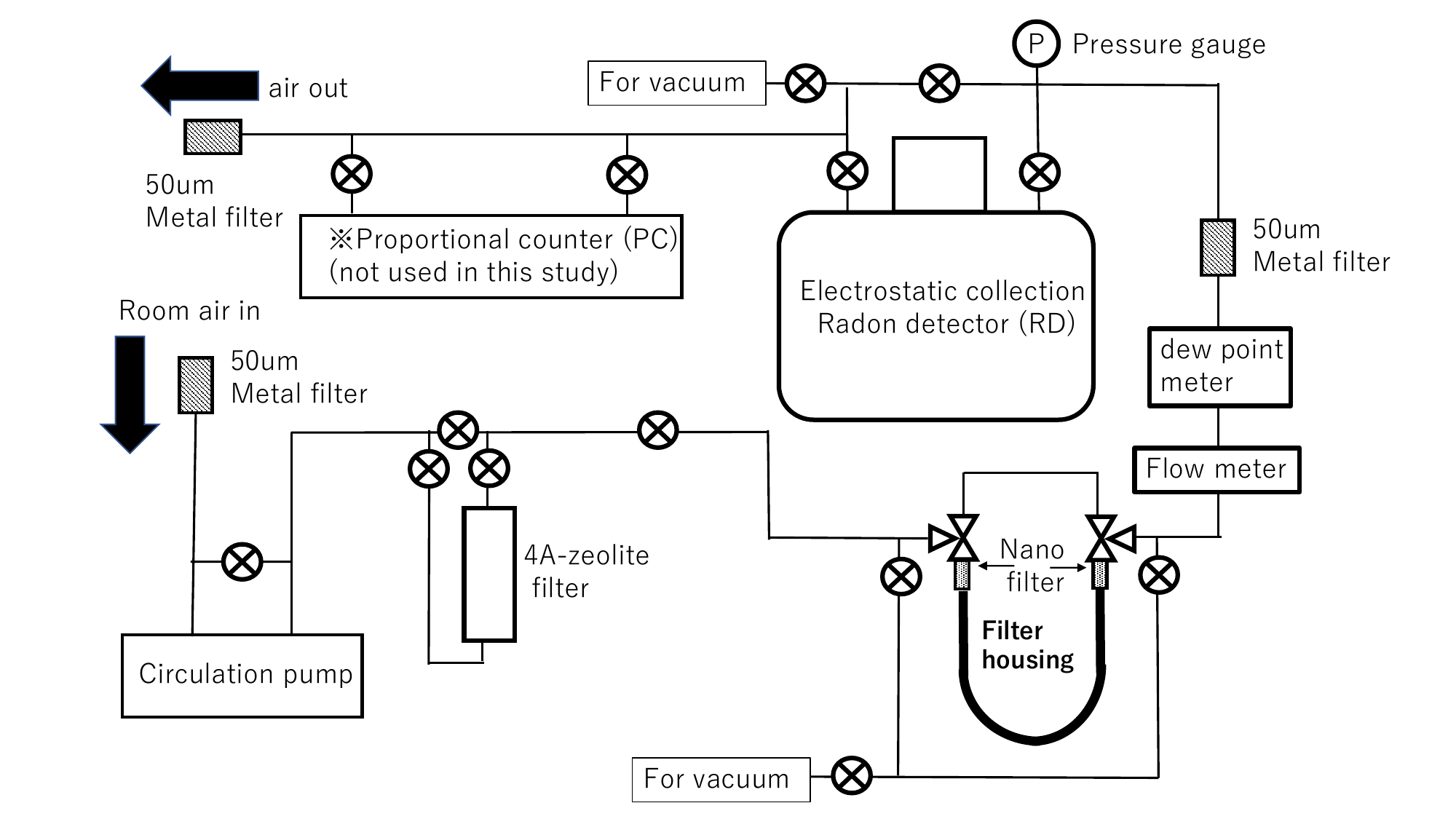}
 \end{center}
  \caption{A schematic representation of the gas line and the radioactivity measurement system for one path system. }
  \label{fig:onepath}
\end{figure}

In an actual air purification system, it is desirable to remove radon in a single pass through a filter after ambient air is drawn into the system. 
Therefore, as an initial test in accordance with the operation of this actual air purification system, the radon removal measurements were conducted with open flow. Figure~\ref{fig:onepath} shows a schematic representation of the gas line and the radioactivity measurement system for open system. Indoor air in the laboratory is taken in by a circulation pump for 0.5~L/min. The air is passed through a 200~g of zeolite 4A-type (4A) filter to remove moisture, then through a 20~g Ag-zeolite sample, that is, Ag-FER8$\%$. Air is released into the room through the RD. For pretreatment of the zeolite, Ag-FER8$\%$ was same treatment with the one of the previous circulation test, and 4A was baked at 200$^{\circ}$C for two hours in a thermostatic chamber.

\begin{figure}[htbp]
  \begin{center}
    \includegraphics[keepaspectratio=true,height=80mm]{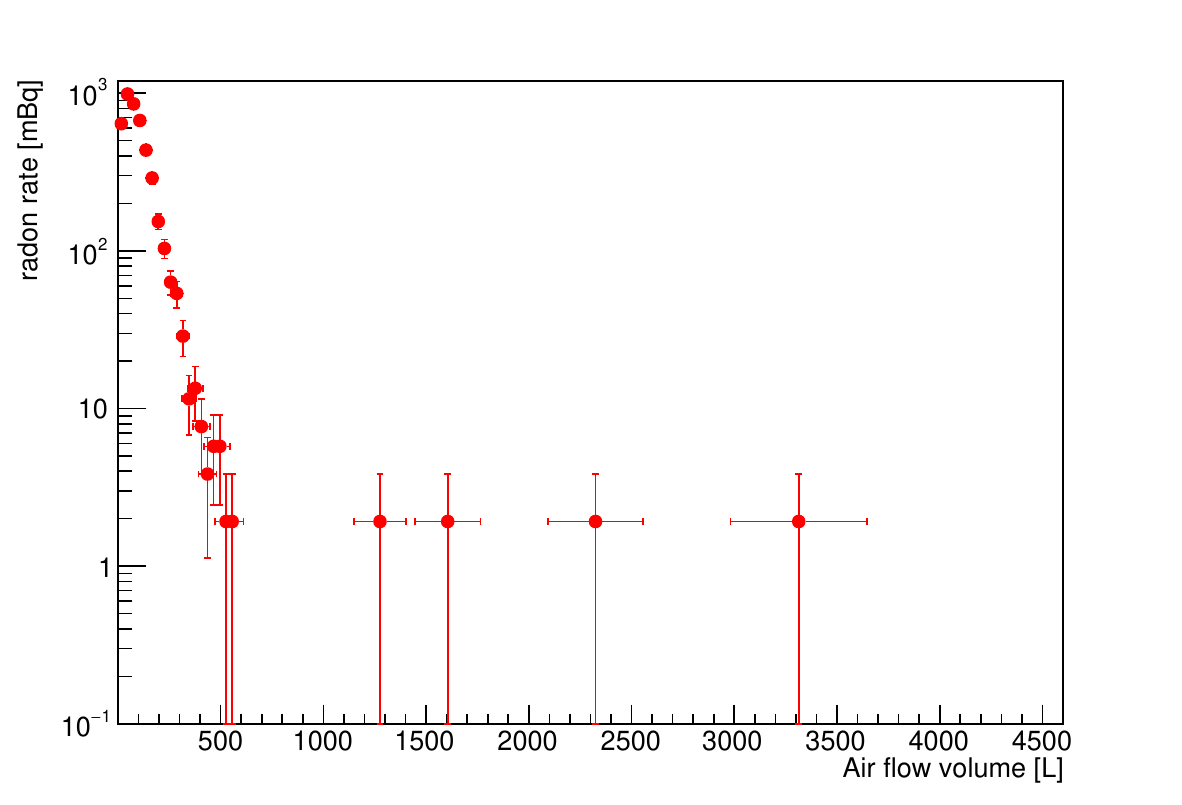}
  \end{center}
  \caption{The time profiles (volume of passing air) of radon reduction factor in one path test.}
  \label{fig:oneptest}
\end{figure}

Figure~\ref{fig:oneptest} shows the passing air volume profiles (the air flow speed 0.5~L/min $\times$ elapsed times to RD) of radon counting rate in one path test. At the beginning of the measurement ($\sim$0~L at Fig.~\ref{fig:oneptest}), ambient air was passed only through 4A, and since 4A has no radon adsorption capacity, the Rn counting rate increased. Then, by passing air through the FH, the radon counting rate decreased rapidly. After one week of measurement, the nanoparticle filter placed upstream of the 4A became clogged with moisture and impurities, then air flow became to zero and the test was stopped. At this point, no increase in radon content was observed, indicating that Ag-FER8$\%$ still has sufficient capacity to adsorb radon. It was found that a 20 g Ag-FER8$\%$ sample was able to remove radon from at least 4,500~L of ambient air in this experiment. The adsorption capacity of radon shows order 10$^{4}$ or more radon removal during the test period.

\subsection{Radon emanation}
\begin{figure}[htbp]
  \begin{center}
    \includegraphics[keepaspectratio=true,height=50mm]{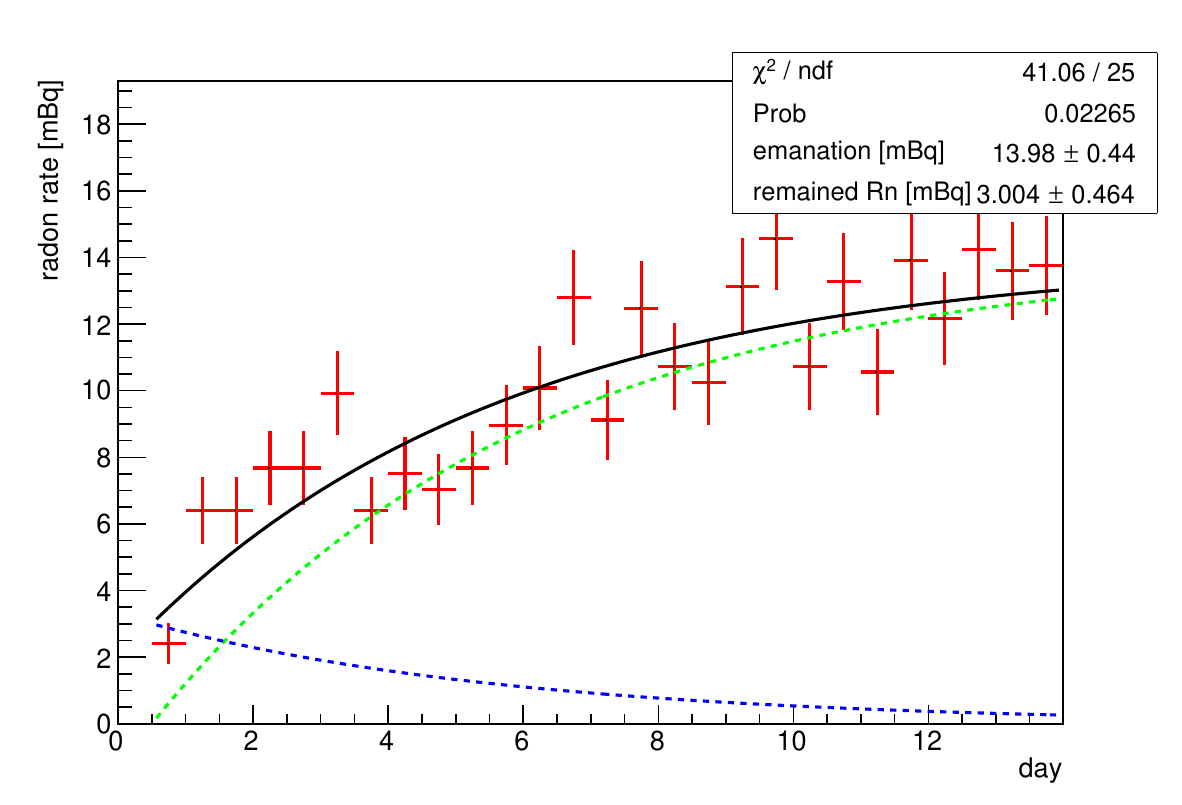}
    \includegraphics[keepaspectratio=true,height=50mm]{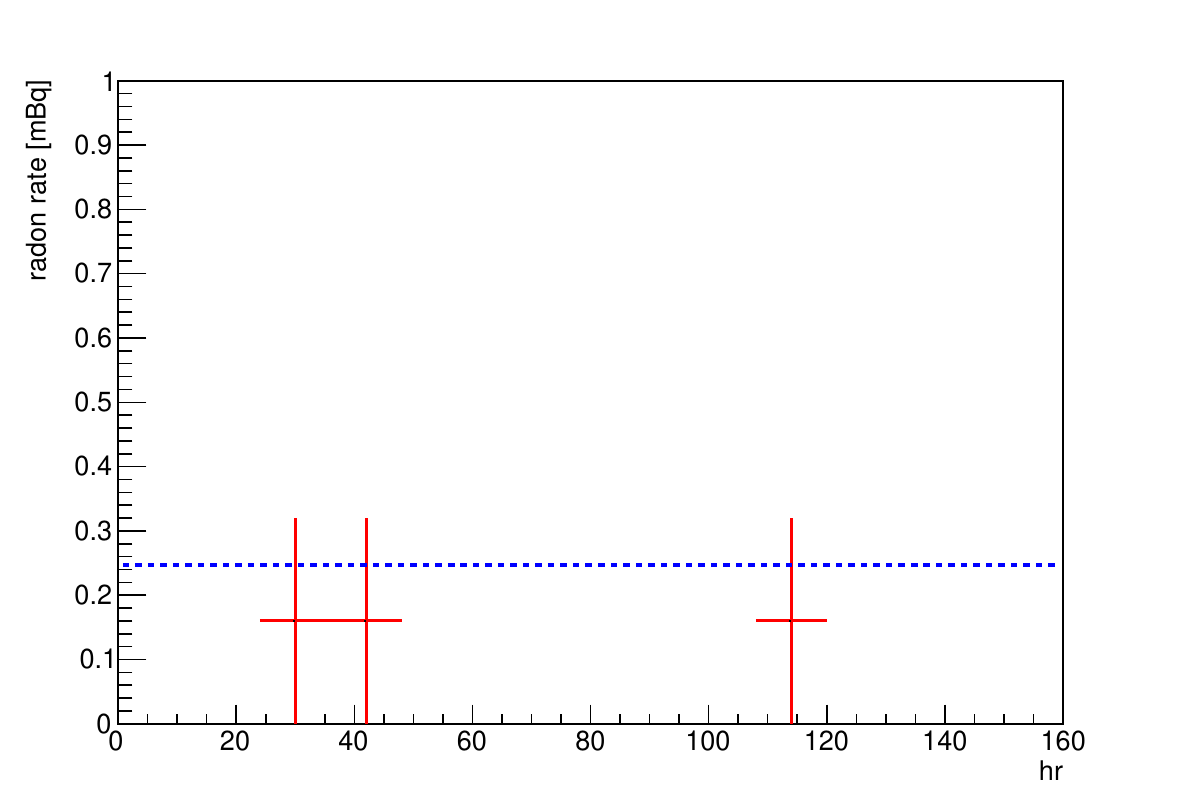}
  \end{center}
  \caption{(Left) The time profile of the radon emanation test with MFI (no Ag ion). The blue dash line indicates the remained radon rate and green dash line indicates the radon emanation rate. (Right) The time profile of the radon emanation test with Ag-MFI. The blue dash line indicates the approximate amount of radon emanation due to upwelling from the system, estimated by a separate measurement.}
  \label{fig:emanation}
\end{figure}

The emanation of radon in Ag-zeolite samples was investigated. As measurement samples, we prepared a MFI 20~g each of sample without an Ag ion exchange and a zeolite with an Ag ion exchange. Radon emanation is evaluated by observing the upwelling of radon over time from zeolite. Measurements of each sample were performed independently. Each samples were placed in a FH, evacuated at 350$^{\circ}$C and circulated with purified G1-grade air in room temperature. Figure~\ref{fig:emanation} shows the time profile of the radon emanation test with MFI (no Ag ion) and Ag-MFI. The time profile data of MFI was fitted as :
\begin{equation}
A(t) = A_{ema}(1-e^{-t/\tau})+A_{BG}e^{-t/\tau}, 
\end{equation}
here $A(t)$ is total radon rate, $A_{ema}$ [mBq] is radon emanation rate and $A_{BG}$ [mBq] is remained radon rate in the RD. $\tau = 3.8235/\ln 2$ [days] is the lifetime of $^{222}$Rn. MFI has large amount of emanation with $A_{ema}\sim$700~mBq/kg (14~mBq/20~g) besides Ag-MFI has no significant emanation, as were the other activated carbon adsorbent candidates~\cite{fiber}. This indicates that MFI itself has impurities containing radium, which produces radon, but in Ag-MFI, no radon is released due to its own ability to adsorb radon. It was determined that the current material, Ag-zeolite, is clean enough less radon emanation which is less than the approximate amount of radon emanation due to upwelling from the system, estimated by a separate measurement, $\sim$0.25 mBq for system. Therefore, a Ag-zeolite made from the same material as the current Ag-zeolite sample shall be used for the air purification system.

\subsection{The adsorption performance of solidified Ag-zeolite}
\begin{figure}[htbp]
  \begin{center}
    \includegraphics[keepaspectratio=true,height=60mm]{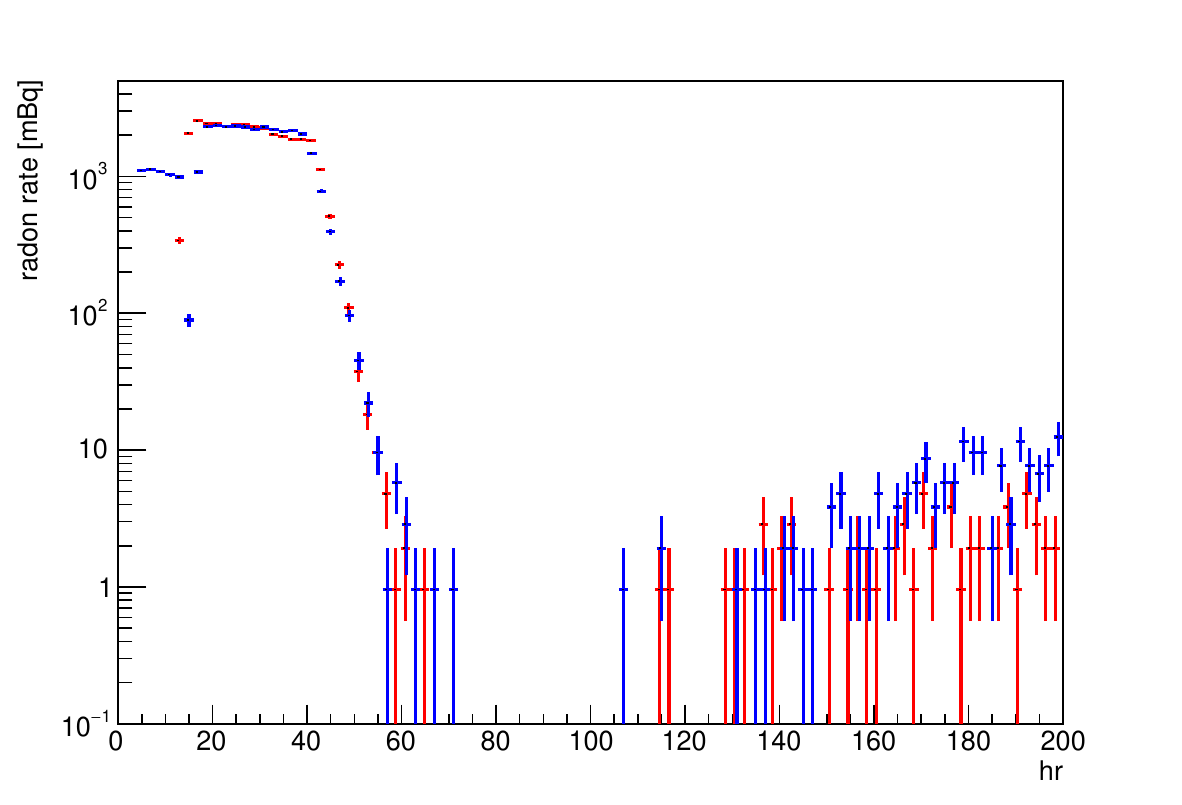}
  \end{center}
  \caption{The time profile of the radon rate during the radon removal test with Ag-MFI sample (red) and solidificated one by PL-7 (blue).}
  \label{fig:solid}
\end{figure}

 The adsorption performance of Ag-zeolite solidified with a solidifier was evaluated. In the adsorption tests in sections~\ref{subsec:close} and~\ref{subsec:open}, Ag-zeolite was solidified as a pressed powder. However, considering the long-term stability of the zeolite filter, solidification with a solidifier is desirable. Therefore, sample Ag-MFI was solidified with 20~$\%$ colloidal silica (FUSO CHEMICAL CO., LTD. PL-7), three times its weight at 350$^{\circ}$C. As a result of the solidification process, the weight of the solidified sample increased by 50$\%$. The solidified Ag-zeolite was crushed to a reasonable size from 0.7 to 2 mm, 30 g was placed in a filter housing, and the same procedure as in section~\ref{subsec:close} was used for radon removal tests.
Figure~\ref{fig:solid} shows the time profile of the radon counting rate during the radon removal test with Ag-MFI sample and solidificated one by PL-7. Although the increase in radon counting rate after RnNO seems significant, no significant difference is found for the removal efficiency and the length of RnNO time. Then, solidification was found not to reduce adsorption efficiency. In the future, we aim to solidify a shaped product in order to guarantee adsorption efficiency in air purification systems.
\subsection{Effect of moisture on adsorption performance}
\label{sec:moisture}
\begin{figure}[htbp]
  \begin{center}
    \includegraphics[keepaspectratio=true,height=80mm]{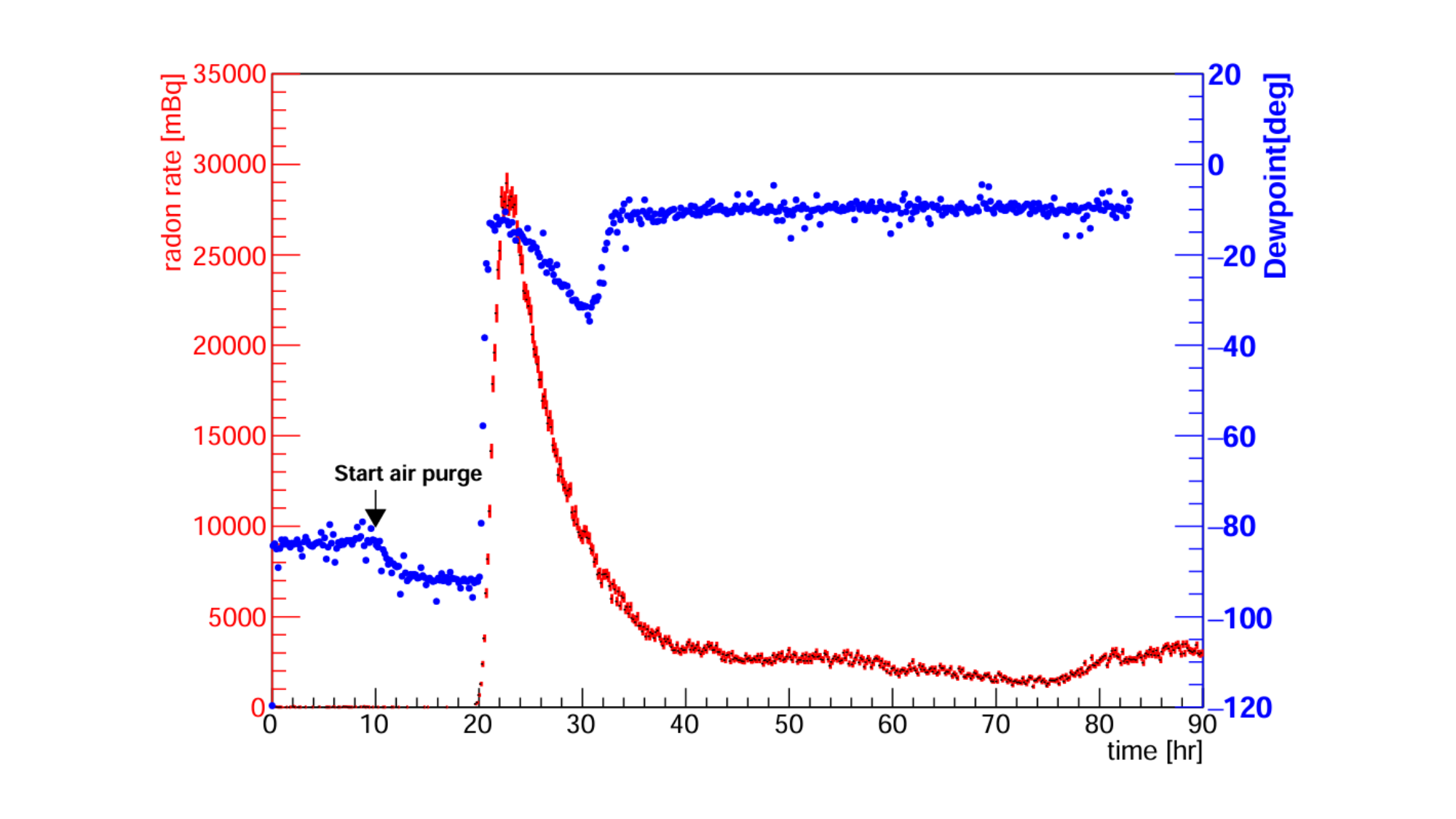}
  \end{center}
  \caption{The time profile of the radon counting rate (red) and dewpoint (blue) during the radon removal test by one pass after nanoparticle filter was removed.}
  \label{fig:failed2}
\end{figure}

Furthermore, the following are the effect of moisture on radon adsorption performance. The radon adsorption test with ambient air was resumed except for the nanoparticle filter which was blocked with impurities in Sec 3.3. Thus, as in Sec 3.3, the ambient air passes through 4A for moisture removal and then through Ag-zeolite. Figure~\ref{fig:failed2} shows the results of a radon removal test after resuming. The dew point decreased immediately after reopening because the 4A was adsorbing moisture. Subsequently, the 4A became saturated and no longer adsorbed moisture, which allowed ambient air containing moisture to enter the Ag-zeolite and release radon.
After this, the dew point increased and decreased once before equilibrium. It is assumed that the radon that had previously been adsorbed is released, and at the same time, moisture is adsorbed onto the Ag-zeolite. The water and radon contents of Ag-zeolite were determined from the time distributions of dew point and radon content in Fig.~\ref{fig:failed2}, respectively. The details are discussed in Appendix A. Then $2\times10^{22}$ of water molecules (equivalent to 0.6~g) and $3\times10^{7}$ of radon atoms were estimated. After Rn was released, the sum of the amount of radon in the ambient air and the amount of radon released from the 4A~\cite{Ogawa1} was observed. Therefore, air purification systems using Ag-zeolite require measures such as removing sufficient moisture and radon from the Ag-zeolite in advance, monitoring the dew point and shutting down the system if moisture is introduced.

\section{Conclusion}
\label{sec:concl}
This paper investigates the removal of radon from purified and ambient airs by Ag-zeolite samples Ag-MFI and Ag-FER. The results show that the Ag-FER sample has order 10$^{4}$ radon removal and has a adsorption capacity for radon than Ag-MFI. Sufficient radon removal capacity was confirmed not only in the case of air circulation, but also in the one path test, in which outside air is directly taken in.
The adsorption capacity of Ag-zeolite for radon does not decrease when solidified with a solidifier. It was also found that there is almost no radon release from the Ag-zeolite itself with order 0.1 mBq. Furthermore, it was found that the incorporation of water either reduced the ability to adsorb radon or rather released radon. These properties will be applied to the Hyper-Kamiokande experiment and to the development of air purification systems for ultra-low radioactivity experiments.

\section*{Acknowledgments}
This work was supported by JSPS KAKENHI Grant Number 23K03435, 19H05807, 24H02243 and the Inter-University Research Program of Institute for Cosmic Ray Research (ICRR), the University of Tokyo.


\appendix
\section{Estimation of the amount of adsorbed water and released radon}
Here are the details of the calculation of the amount of adsorbed water molecules and the amount of radon released in the Sec.~\ref{sec:moisture} measurement.
\subsection*{Amount of adsorbed water molecules}
As shown in Fig.~\ref{fig:failed2}, immediately after moisture incorporation, the measured dew point of the dew point meter downstream of the Ag-zeolite once dropped and then reached equilibrium. The amount of moisture adsorbed during this period was derived by subtracting the amount of moisture determined from the dew point at each measurement point during this period from the average amount of moisture determined from the dew point at equilibrium over 40~hour in Fig.~\ref{fig:failed2}. The interval between each measurement point is 10 minutes, and the air flow rate at each measurement point is 5~L/10~min. The dew point to moisture content was converted using following table~\cite{dewpoint}. The total amount of water adsorbed is determined to be 0.6~g, which corresponds to 2$\times$10$^{22}$ water molecules.
\subsection*{Amount of released radon}
For the conversion of the amount of radon released from Ag-zeolite, it is necessary to create a model that results in the time distribution of the amount of radon in Fig.~\ref{fig:failed2} and derive it from there. The following two factors are taken into account. 

First, the time distribution of radon released from Ag-zeolite, $A(t)$, is defined as,
\begin{equation}
A(t)=A_{0}e^{-(t-t0)/a},
\label{eq-a1}
\end{equation}
where $A_{0}$ is the amount of radon activity immediately after the release and $a$ is the time constant of the release. $t_{0}$ is the start time of Rn release in Fig.~\ref{fig:failed2} and is set to 19.6 hour. Next, the radon detector is 80 L, and air enters the detector passing Ag-zeolite at 0.5 L/min and is released at 0.5 L/min to the outside. Let $C(t)$ be the activity of radon in the RD. Here the time variation of $C(t)$ due to air entering and leaving the detector is,
\begin{equation}
\frac{dC(t)}{dt}=A(t)+B-\gamma C(t)
\label{eq-a2}
\end{equation}
where $B$ is the amount of radon activity from the external air, assumed constant and obtained from the average of radon activity at $t=50-58$~hour in Fig.~\ref{fig:failed2}. $\gamma$ is the rate of release from the detector to the outside, with $\gamma=vdt/V$ ($v$: air flow rate 0.5~L/min, $V$: 80L).

\begin{figure}[htbp]
  \begin{center}
    \includegraphics[keepaspectratio=true,height=80mm]{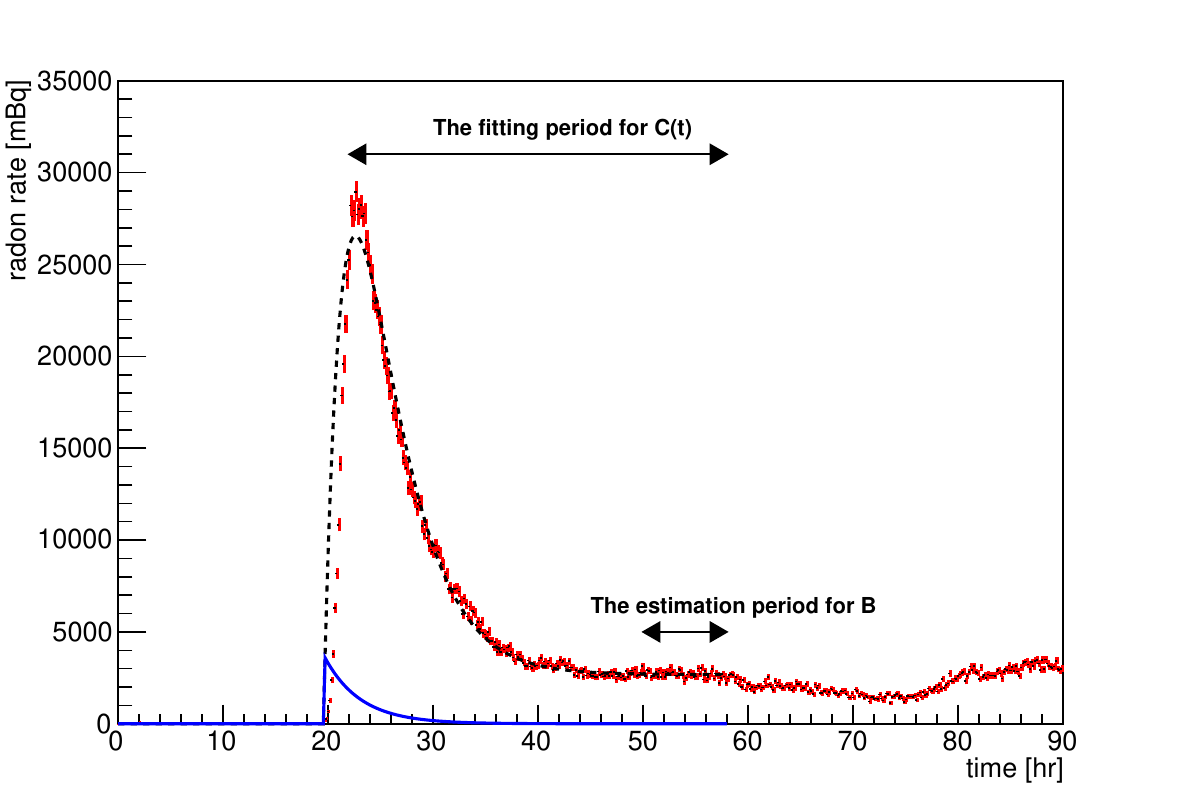}
  \end{center}
  \caption{The time profile of the radon counting rate (red). The assumed $A(t)$ (blue-solid line) and $C(t)$ (black-dashed line) for the best fit are also shown. The two arrows point to the fitting period for $C(t)$ and the estimation period for $B$, respectively.}
  \label{fig:Rnest}
\end{figure}

In order to reproduce the radon release measurement data ($t=22-58$~hour) in Fig.~\ref{fig:failed2}, $A_{0}$ and a in Eq.~\ref{eq-a1} are derived by fitting the time distributions of radon activity variables obtained from the model considering two factors, $A_{0}$ and $a$. The resulting graph is shown in Fig.~\ref{fig:Rnest}. The derived $A(t)$ and $C(t)$ are shown here. To simplify the calculation of the estimates, we do not consider sequential equilibria of $^{222}$Rn-$^{214}$Po for RD. The discrepancy between the measured data and the calculation at the rise around 20~hr may be attributed to this. From this result, it can be seen that $A_{0}=3590$(mBq) and $a=3.4$(hour). The number of released radon atoms $D$ is calculated by
\begin{equation}
D=\int{A(t)}dt/\lambda
\label{eq-a3}
\end{equation}
where $\lambda$ is decay coefficient for $^{222}$Rn. As the result, $D=3\times10^{7}$ was estimated.


\begin{thebibliography}{00}
\bibitem{Ianni} A. Ianni, SciPost Phys. Proc. 12, 007 (2023) 
\bibitem{NIM-SK} Y. Nakano et. al, Nucl.Instrum.Meth.A997:164297,2020
\bibitem{SK-solar} K. Abe et. al (Super-Kamiokande Collaboration), Phys. Rev. D 109, 092001,2024  
\bibitem{Fukuda} S.  Fukuda et. al (Super-Kamiokande Collaboration) Nucl.Instrum.Meth.A501:418-462,2003
\bibitem{HKdesign} K. Abe et. al (Hyper-Kamiokande proto-collaboration), arXiv:1805.04163
\bibitem{SKair} Y. Takeuchi et. al, Phys. Lett., B452:418–424, 1999. 
\bibitem{Nakano} Y. Nakano et. al, Nucl.Instrum.Meth.A867:108-114,2017
\bibitem{Fukui} M. Fukui et al., Xe adsorption performance of ag-loaded zeolite, In The 36th Zeolite Research Presentation (on-line). The Zeolite Institute of Japan (2020).
\bibitem{SHeinitz} S. Heinitz et. al, Nature Scientific Reports vol.13, 6811 (2023)
\bibitem{Oleksandra} O. Veselska et. al, Prog. Theor. Exp. Phys. 2024 023C01
\bibitem{Ogawa1} H. Ogawa et al., J.Instrum. \textbf{19} P02004 (2024). 
\bibitem{Hosokawa} K. Hosokawa et. al, Prog. Theor. Exp. Phys. 2015, 033H01 
\bibitem{fiber} Y. Nakano et. al, Prog. Theor. Exp. Phys. 2020, 113H01
\bibitem{dewpoint} https://official.koganei.co.jp/common/pdf/tech/E4041\underline{ }volume\underline{ }rate\underline{ }conversion.pdf
\end{thebibliography}
\end{document}